\documentclass[12pt,a4paper]{article}
\usepackage{graphicx}
\usepackage{amssymb}
\usepackage{amsmath}
\usepackage{bm}
\usepackage{color}
\usepackage{tikz}
\usepackage{theorem}
\usepackage{cite}
\usepackage{amsfonts}
\usepackage{multirow}
\usepackage{changepage}
\usepackage{booktabs}
\usepackage{slashed}
\usepackage{graphicx}
\usepackage{caption}
\usepackage{subcaption}
\usepackage{nicematrix}
\usepackage{physics}
\usepackage{geometry}
\usepackage[compat=1.0.0]{tikz-feynman}

\usepackage[sort&compress,numbers, merge]{natbib}

\newcommand{\eprint}[2][]{\href{https://arxiv.org/abs/#2}{arXiv:~\nolinkurl{#2}}}

\usepackage{scrextend}

\newcommand{\sabk}[1]{\left[{#1}\right\rangle}
\newcommand{\asbk}[1]{\left\langle{#1}\right]}

\definecolor{Orange}{cmyk}{0,0.61,0.87,0}
\definecolor{JungleGreen}{cmyk}{0.99,0,0.52,0} 
\definecolor{OliveGreen}{cmyk}{0.64,0,0.95,0.40}
\definecolor{Brown}{cmyk}{0,0.81,1,0.60}
\definecolor{RoyalBlue}{cmyk}{0.71,0.53,0,0.12}
\definecolor{darkspringgreen}{rgb}{0.09, 0.45, 0.27}

\newcommand{\be}{\begin{equation}}
\newcommand{\ee}{\end{equation}}
\newcommand{\bea}{\begin{eqnarray}}
\newcommand{\eea}{\end{eqnarray}}

\usetikzlibrary{decorations.markings,angles,quotes,positioning,decorations.pathreplacing,decorations.markings,snakes,arrows,calc}
\usetikzlibrary{fadings}
\usetikzlibrary{shapes.misc}
\tikzset{crossr/.style={cross out, draw=red, minimum size=4*(#1-\pgflinewidth), inner sep=0pt, outer sep=0pt},
crossr/.default={2pt}}
\tikzset{crossb/.style={cross out, draw=black, minimum size=4*(#1-\pgflinewidth), inner sep=0pt, outer sep=0pt},
crossb/.default={2pt}}
\tikzset{crossp/.style={cross out, draw=violet, minimum size=4*(#1-\pgflinewidth), inner sep=0pt, outer sep=0pt},
crossp/.default={2pt}}

\tikzfading 
[
  name=fade out,
  inner color=transparent!0,
  outer color=transparent!100
]

\tikzset{
upp/.style={postaction={decorate},
  decoration={markings,mark=at position .5 with  \arrow{>}}},
  dow/.style={postaction={decorate},
  decoration={markings,mark=at position .5 with  \arrow{<}}},
 }

\tikzset{
    ncbar angle/.initial=90,
    ncbar/.style={
        to path=(\tikztostart)
        -- ($(\tikztostart)!#1!\pgfkeysvalueof{/tikz/ncbar angle}:(\tikztotarget)$)
        -- ($(\tikztotarget)!($(\tikztostart)!#1!\pgfkeysvalueof{/tikz/ncbar angle}:(\tikztotarget)$)!\pgfkeysvalueof{/tikz/ncbar angle}:(\tikztostart)$)
        -- (\tikztotarget)
    },
    ncbar/.default=0.5cm,
}

\tikzset{round left paren/.style={ncbar=0.5cm,out=100,in=-100}}
\tikzset{round right paren/.style={ncbar=0.5cm,out=80,in=-80}}

\usepackage[colorlinks=True, citecolor=blue, linkcolor=blue, urlcolor=black]{hyperref}

\allowdisplaybreaks[1]

\begin{document}

\begin{titlepage}
\begin{center}
\hfill

\vspace{2.0cm}
{\Large\bf  
Entanglement Maximization and Symmetry Selection in Composite Higgs Models
}

\vspace{2cm}
{\bf 
Cihang Li$^1$, Teng Ma$^{2,3}$, Jing Shu$^{4,5,6}$ and Mingdi Zhu$^{2,3}$
}
\\
\vspace{0.7cm}
{\it\footnotesize

${}^1$School of Physics, Peking University, Beijing, 100871, China\\
${}^2$International Center for Theoretical Physics Asia-Pacific (ICTP-AP),\\ University of Chinese Academy of Sciences (UCAS), Beijing, 100190, China\\
${}^3$Taiji Laboratory for Gravitational Wave Universe (Beijing/Hangzhou),\\ University of Chinese Academy of Sciences (UCAS), Hangzhou, Zhejiang, 310024, China\\
${}^4$School of Physics and State Key Laboratory of Nuclear Physics and Technology,\\ Peking University, Beijing, 100871, China\\
${}^5$Center for High Energy Physics, Peking University, Beijing, 100871, China\\
${}^6$Beijing Laser Acceleration Innovation Center, Huairou, Beijing, 101400, China\\

}
\vspace{0.9cm}

\abstract
\par Recent developments suggest that the extremization of quantum entanglement may provide a useful organizing principle for strong dynamics. While entanglement suppression characterizes low-energy QCD, we investigate the role of entanglement maximization in the electroweak symmetry breaking sector. Focusing on the Composite Higgs Model, we analyze the process $hh \to t\bar{t}$ by treating the fermionic helicity space as a bipartite quantum system. Maximal entanglement imposes nontrivial constraints on the fermionic effective theory and leads to two simple symmetry structures in the top sector. One is the Maximal Symmetry branch, characterized by the vanishing of the Higgs-dependent form factor $\Pi_1$ and the finiteness of the Higgs potential. The other is a generalized $Z_2$-matching branch relating the left- and right-handed top sectors. Our results establish a quantitative connection between entanglement structure and the naturalness of electroweak symmetry breaking, and suggest that the symmetry patterns of the strong sector may be understood from the perspective of entanglement extremization.

\end{center}
\end{titlepage}
\setcounter{footnote}{0}

\section{Introduction}
\par Symmetry principles constitute a foundational element of modern theoretical physics, delineating the fundamental interactions within the Standard Model and guiding the construction of Effective Field Theories. While local gauge symmetries provide a rigid structure for electroweak and strong interactions, the origin of global symmetries remains a profound conceptual puzzle. These symmetries are frequently introduced as axiomatic inputs at high-energy scales or treated as accidental features of the low-energy spectrum, particularly when invoked to address the hierarchy problem and protect scalar masses from radiative corrections~\cite{tHooft:1979rat,Weinberg:1978kz}. However, the rapidly evolving intersection of high-energy physics and quantum information science suggests that these structural properties of quantum field theories may not be fundamental postulates but rather emergent phenomena arising from deeper information-theoretic constraints~\cite{Wheeler1990Information}.

\par The incorporation of quantum information theory into high-energy physics extends beyond the philosophical ``It from Bit'' proposition to offer concrete analytical tools for theoretical investigation. This interdisciplinary approach has progressively developed into a quantitative framework wherein concepts such as entanglement entropy and quantum circuit complexity contribute to our understanding of the emergence of spacetime geometry~\cite{Ryu_2006,Maldacena_2013,VanRaamsdonk:2010pw,Pasquale_Calabrese_2004,Maldacena:2001kr}. In this context, the S-matrix governing particle scattering may be analyzed as a unitary quantum gate operating on the Hilbert space of internal quantum numbers~\cite{low2021symmetry,Carena:2023vjc,Preskill:2018fag,Jordan_2012}. By viewing fundamental interactions through the lens of quantum logic operations, the generation or modification of quantum entanglement offers a novel means to characterize the underlying dynamics of the theory. Furthermore, as the entangling capacity of unitary operations is well characterized in quantum information theory~\cite{kraus2001optimal}, this correspondence provides a pathway to mapping scattering amplitudes onto the language of quantum computation. More broadly, this perspective suggests that the extremization of entanglement may encode structural information about strongly coupled dynamics and their emergent symmetries~\cite{low2021symmetry,Cervera_Lierta_2017,Beane_2019}.

\par A significant body of recent research investigates the quantitative connection between symmetry emergence and the behavior of quantum entanglement in scattering amplitudes. Pioneering studies in low-energy quantum chromodynamics revealed a striking correlation between the suppression of spin entanglement in nucleon-nucleon scattering and the emergence of Wigner's $SU(4)$ spin-flavor symmetry~\cite{Beane_2019,Liu_2023}. This discovery led to the conjecture that dynamical entanglement suppression is a key feature of strong interactions in the infrared. While these investigations highlight systems that act as minimal entanglers, other recent analyses have explored the implications of entanglement maximization~\cite{carena2025entanglement}, suggesting that specific maximal-entanglement configurations, analogous to the Bell states observed in QED processes~\cite{Cervera_Lierta_2017}, can also act as attractors for symmetric vacua. Collectively, these findings indicate that entanglement extremization may provide a useful diagnostic of enhanced symmetry, both in systems that suppress quantum correlations and in those that maximize them.

\par This information-theoretic perspective offers a novel vantage point for addressing the naturalness of the electroweak scale, which remains a significant theoretical puzzle within the Standard Model. A compelling solution is provided by Composite Higgs theory~\cite{Kaplan:1983sm,Georgi:1984af}, which posits that the Higgs doublet consists of pseudo-Nambu-Goldstone bosons (pNGBs) arising from a new strong dynamics, conceptually analogous to pions in QCD. In this framework, the pNGB Higgs potential is radiatively generated by gauge boson and top quark loops that explicitly break the global symmetry of the composite sector. However, this potential is generically divergent and sensitive to the confinement scale of the strong dynamics. Consequently, explaining the vast hierarchy between the confinement and electroweak scales without excessive fine-tuning is challenging. This specific theoretical difficulty is known as the Little Hierarchy problem. Maintaining this separation typically requires the introduction of additional structural protections, such as maximal symmetry~\cite{Csaki:2017cep,Csaki:2018zzf,Csaki:2019coc} or collective symmetry breaking~\cite{Arkani-Hamed:2001nha}. Although these mechanisms ensure a finite potential, they are often implemented as heuristic postulates. Fundamentally, a deeper dynamical justification is currently missing. It remains an open question whether there exists a more general principle that drives the strong sector toward these special symmetric configurations.

\par In this work, we investigate the structural constraints imposed by entanglement maximization on strong dynamics. We apply this information-theoretic principle to electroweak symmetry breaking (EWSB) in the Minimal Composite Higgs Model (MCHM) based on the $SO(5)/SO(4)$ coset space~\cite{Agashe:2004rs,marzocca2012general}. Within this framework, the top quark serves as a unique probe for this analysis. As the heaviest particle in the Standard Model, it possesses the strongest coupling to the electroweak symmetry breaking sector. Moreover, its ultra-short lifetime ensures that spin information is preserved in the decay products. This property renders the top-antitop pair ($t\bar{t}$) an ideal laboratory for analyzing quantum correlations. While previous studies have largely treated entanglement as a diagnostic of symmetry emergence, we instead impose the condition of maximal quantum correlation directly on the fermionic top sector and derive the resulting effective field theory (EFT) constraints. We focus on the scattering of Higgs bosons into top-antitop quark pairs ($hh \to t\bar{t}$). By mapping the helicity space of the final-state fermions onto a bipartite quantum system, we quantify the entanglement using concurrence~\cite{PhysRevLett.80.2245,Hill_1997}. We then constrain the effective Lagrangian by requiring the scattering amplitude to generate a maximally entangled final state.

\par This investigation establishes a quantitative link between the entanglement properties of the fermion sector and the naturalness of the Higgs potential. Our analysis shows that the requirement of maximal entanglement imposes nontrivial restrictions on the effective description of the strong dynamics. Rather than selecting an isolated parameter point, this condition naturally leads to two simple structural ways of satisfying the entanglement constraint in the top sector. One is the Maximal Symmetry branch~\cite{Csaki:2017cep}, characterized by the vanishing of the Higgs-dependent form factor $\Pi_1$ and by the finiteness of the Higgs potential, which is central to reducing electroweak tuning. The other is a generalized $Z_2$-matching branch, in which the left- and right-handed top sectors obey a nontrivial relation among the relevant form-factor combinations. In this way, the information-theoretic constraint provides a new perspective on the symmetry structures associated with the naturalness of EWSB.

\par The remainder of this paper is organized as follows. In Sec.~\ref{sec:Ent}, we establish the kinematic framework by mapping the top quark helicity space onto a bipartite quantum system and defining concurrence as the measure of entanglement. Sec.~\ref{sec:CHM} reviews the Minimal Composite Higgs Model and the concept of Maximal Symmetry as a solution to the hierarchy problem. The core analysis is presented in Sec.~\ref{Sec:EC}, where we derive the constraints on the effective Lagrangian imposed by entanglement maximization in $hh\to t\bar t$ scattering. We also briefly contrast the result with the gauge sector, where the relevant amplitude matching follows automatically from the covariant-derivative structure. Technical details of this comparison are collected in Appendix B. Finally, we present our concluding remarks and future outlook in Sec.~\ref{Sec:Concl}.

\section{Helicity Amplitudes and Quantum Entanglement}
\label{sec:Ent}

\par To investigate the emergent symmetries in the composite sector, we analyze the kinematic structure of the top-antitop quark pairs produced in Higgs boson scattering through the lens of quantum information theory. Recent developments have established that the top quark, owing to its exceptionally short lifetime, decays before hadronization and thus preserves its spin information. This property renders the $t \bar{t}$ system an ideal laboratory for probing quantum correlations at high energy scales~\cite{Afik_2021,Fabbrichesi_2021}. The fundamental observables describing the internal degrees of freedom of these final-state fermions are their helicities. We focus on the symmetric phase in which the composite Higgs field has not yet acquired a vacuum expectation value, corresponding to the high-energy regime where the scattering energy significantly exceeds the top quark mass. This ensures that chirality and helicity coincide, rendering helicity a well-defined quantum number for the bipartite analysis.

\par Since these helicity eigenstates form a complete orthonormal basis, they admit a direct mapping of the scattering products onto a bipartite quantum system~\cite{Aguilar_Saavedra_2022}. By identifying the left-handed and right-handed helicity states of the top and antitop quarks with the computational basis states of a qubit, we define
\begin{equation}
\ket{t_L} \equiv \ket{0}, \qquad \ket{t_R} \equiv \ket{1}.
\end{equation}
This identification maps the combined Hilbert space of the final-state fermions, $\mathbb{C}^2 \otimes \mathbb{C}^2$, directly onto the logical basis. Applying the same mapping to the antitop sector, the four physical product states correspond to
\begin{equation}
\ket{t_L\bar{t}_L} \to \ket{00}, \qquad
\ket{t_L\bar{t}_R} \to \ket{01}, \qquad
\ket{t_R\bar{t}_L} \to \ket{10}, \qquad
\ket{t_R\bar{t}_R} \to \ket{11}.
\end{equation}

\par Within this kinematic framework, the entanglement properties of the scattering process $hh \to t \bar{t}$ are encoded in the helicity amplitudes $\mathcal{M}_{\lambda \bar{\lambda}}$, where $\lambda, \bar{\lambda} \in \{L,R\}$ denote the helicities of the top and antitop quarks. Since the initial state is a color singlet, the color structure plays no role in the entanglement analysis. The spin-helicity configuration at fixed kinematics may therefore be described by a pure quantum state $\ket{\Psi}$~\cite{Cervera_Lierta_2017,Fabbrichesi_2021}, constructed as
\begin{equation}
\label{eq:psi_state}
\ket{\Psi} = \frac{1}{\sqrt{\mathcal{N}}}
\left(
\mathcal{M}_{LL}\ket{00}
+\mathcal{M}_{LR}\ket{01}
+\mathcal{M}_{RL}\ket{10}
+\mathcal{M}_{RR}\ket{11}
\right),
\end{equation}
where
\begin{equation}
\mathcal{N} = |\mathcal{M}_{LL}|^2 + |\mathcal{M}_{LR}|^2 + |\mathcal{M}_{RL}|^2 + |\mathcal{M}_{RR}|^2
\end{equation}
is the normalization factor proportional to the differential cross-section. This establishes a direct correspondence between the scattering amplitudes and the state vectors used in quantum information analysis~\cite{low2021symmetry}.

\par To quantify the quantum correlations encoded in this state, we use the concurrence ($\Delta$), a standard measure of entanglement that is monotonic with the entanglement of formation~\cite{PhysRevLett.80.2245}. While the general definition for mixed states involves diagonalizing a spin-flipped density matrix~\cite{Hill_1997}, for a pure state of the form given in Eq.~(\ref{eq:psi_state}) the concurrence reduces to a simple algebraic expression depending only on the helicity amplitudes. It is given by twice the magnitude of the determinant of the corresponding $2\times 2$ coefficient matrix,
\begin{equation}
\label{eq:concurrence}
\Delta(\ket{\Psi}) = \frac{2}{\mathcal{N}}
\left|
\mathcal{M}_{LL}\mathcal{M}_{RR}
-
\mathcal{M}_{LR}\mathcal{M}_{RL}
\right|.
\end{equation}
The value of $\Delta$ ranges from $0$ for a separable state to $1$ for a maximally entangled state. This form is particularly useful for our purposes, since it translates the condition of maximal entanglement directly into constraints on the scattering amplitudes.

\par In the Composite Higgs framework, we take maximal entanglement as a working criterion and examine the consequences of imposing it on the final state. Concretely, we require the concurrence to satisfy
\begin{equation}
\Delta = 1,
\end{equation}
so that the scattering amplitude generates a maximally entangled state, analogous to a Bell-state configuration such as $(\ket{00}+\ket{11})/\sqrt{2}$~\cite{Cervera_Lierta_2017}. As we shall see, imposing this condition on the S-matrix elements, namely
\begin{equation}
\left|
\mathcal{M}_{LL}\mathcal{M}_{RR}
-
\mathcal{M}_{LR}\mathcal{M}_{RL}
\right|
=
\frac{\mathcal{N}}{2},
\end{equation}
leads to nontrivial constraints on the form factors of the effective Lagrangian~\cite{Csaki:2017cep,Barr_2022}. This framework will be used in the following sections to derive the corresponding EFT constraints and their symmetry implications.

\section{Basics of the Composite Higgs Model and Maximal Symmetry}
\label{sec:CHM}

\par To provide a physical context for our information-theoretic analysis, we adopt the framework of the MCHM. The underlying symmetry breaking pattern is characterized by the coset space $SO(5)/SO(4)$, wherein the associated pNGB Higgs fields $h^{\hat a}$ are described by the non-linear sigma field
\begin{equation}
U = \exp\left(i \frac{\sqrt{2}}{f} h^{\hat{a}} T^{\hat{a}} \right),
\end{equation}
where $T^{\hat a}$ denote the broken generators of $SO(5)$ and $f$ is related to the compositeness scale of the new strong dynamics~\cite{Agashe:2004rs}. Under the global symmetry group, the field $U$ transforms non-linearly as $U \to g U h^\dagger(h^{\hat{a}},g)$, with $g$ an element of the global group $SO(5)$ and $h$ an element of the unbroken subgroup $SO(4)$.

\par The coupling of Standard Model fermions to the strong sector is realized through the partial compositeness mechanism~\cite{Kaplan:1991dc}, wherein elementary fields mix linearly with composite partners from the strong dynamics to acquire mass. This mixing is the primary source of explicit global symmetry breaking, generating the Higgs potential at loop level and triggering electroweak symmetry breaking via the Coleman-Weinberg mechanism~\cite{PhysRevD.7.1888}. We assume that the elementary top quark doublet and singlet couple to composite fermionic operators transforming in the fundamental representation of $SO(5)$. Following the Callan-Coleman-Wess-Zumino (CCWZ) construction~\cite{Coleman:1969sm,Callan:1969sn}, the relevant effective Lagrangian is
\begin{align}\label{eq:SO5mixing1}
\mathcal{L}_{f} =&\; \bar{\Psi}_Q (i \slashed{\nabla} - M_Q ) \Psi_Q + \bar{\Psi}_S (i \slashed{\nabla} - M_S ) \Psi_S \nonumber \\
&+ \frac{f}{\sqrt{2}} \bar{\Psi}_{t_R} P_L (\epsilon_{tS} U \Psi_{S} + \epsilon_{tQ} U \Psi_{Q}) \nonumber \\
&+ f \bar{\Psi}_{q_L} P_R (\epsilon_{qS} U \Psi_S + \epsilon_{qQ} U \Psi_Q) + \text{h.c.},
\end{align}
where $\epsilon$ denote the mixing parameters, and $\nabla \equiv \partial_\mu - i E_\mu$ is the covariant derivative of the composite fermion, with $E_\mu$ being the CCWZ connection. In this approach, the elementary fields $q_L$ and $t_R$ are embedded into incomplete multiplets within the fundamental representation ${\bf 5}$ of $SO(5)$ as
\begin{equation}
\Psi_{q_L} = \frac{1}{\sqrt{2}}
\begin{pmatrix}
b_L \\
-i b_L \\
t_L \\
i t_L \\
0
\end{pmatrix},
\qquad
\Psi_{t_R} =
\begin{pmatrix}
0 \\
0 \\
0 \\
0 \\
t_R
\end{pmatrix}.
\end{equation}
Furthermore, the composite resonances $\Psi_{Q,S}$ are the top partners in the $\bf 4$ and $\bf 1$ representations of the unbroken symmetry $SO(4)$, with explicit forms
\begin{equation}
\Psi_Q = \frac{1}{\sqrt{2}}
\begin{pmatrix}
i B - i X_{5/3} \\
B + X_{5/3} \\
i T + i X_{2/3} \\
-T + X_{2/3} \\
0
\end{pmatrix},
\qquad
\Psi_S =
\begin{pmatrix}
0 \\
0 \\
0 \\
0 \\
T_1
\end{pmatrix}.
\end{equation}

\par Integrating out the heavy composite resonances $\Psi_{Q,S}$ yields the low-energy effective Lagrangian in momentum space. The effective interactions between the elementary fields and the strong sector can be written as
\begin{equation}
\label{eq:Lagrangian}
\begin{aligned}
\mathcal{L}_{\text{eff}} = &\,
\bar{\Psi}_{q_L} \slashed{p} \bigl[ \Pi_0^q(p) + \Pi_1^q(p) \Sigma^{\prime} \bigr] \Psi_{q_L}
+ \bar{\Psi}_{t_R} \slashed{p} \bigl[ \Pi_0^t(p) + \Pi_1^t(p) \Sigma^{\prime} \bigr] \Psi_{t_R} \\
&+ \bar{\Psi}_{q_L} \bigl[ M_1^t(p) \Sigma^{\prime} \bigr] \Psi_{t_R} + \text{h.c.}
\end{aligned}
\end{equation}
Here, the operator $\Sigma^{\prime}$ is defined as the linearly realized field $\Sigma^{\prime} = U V U^\dagger = U^2 V$, where $V = \text{diag}(1,1,1,1,-1)$ represents the vacuum expectation value breaking the global symmetry. Under the global group, $\Sigma^{\prime}$ transforms as $\Sigma^{\prime} \to g \Sigma^{\prime} g^\dagger$. The scalar functions $\Pi_{0,1}^{q,t}(p)$ denote the momentum-dependent form factors parameterizing the kinetic structure, while $M_1^t(p)$ characterizes the mass-generating term arising from partial compositeness.

\par The asymptotic behavior of these form factors is closely tied to the naturalness of the electroweak scale. Based on dimensional analysis, the form factor $M_1^t$ typically yields a finite contribution to the Higgs potential, scaling as $M_1^t \sim f^2 \epsilon_{q}\epsilon_{t}$. By contrast, the form factor $\Pi_1$ often exhibits a slow decay at large momenta, leading to a Higgs potential that is quadratically divergent. This sensitivity to the ultraviolet (UV) cutoff $\Lambda$ of the strong dynamics exacerbates the Little Hierarchy problem, necessitating substantial fine-tuning to reproduce the observed Higgs mass~\cite{Barbieri:1987fn,Matsedonskyi_2013}. In particular, a non-vanishing $\Pi_1$ implies that vacuum alignment remains sensitive to details of the strong sector at the scale $\Lambda$. Controlling the magnitude of the form factor $\Pi_1^{q,t}$ is therefore essential for constructing a natural theory of electroweak symmetry breaking.

\par It has been shown that an enhanced global symmetry structure, referred to as Maximal Symmetry~\cite{Csaki:2017cep}, provides a mechanism to eliminate the divergence-inducing form factor $\Pi_1^{q,t}$ and reduce the tuning of the Higgs potential. In this framework, the composite sector possesses an enhanced global symmetry that is preserved by the mixing with elementary fields. Within the MCHM, this structure removes the most dangerous radiative contributions to the Higgs potential. In practice, Maximal Symmetry enforces the vanishing of the Higgs-dependent form factors,
\begin{equation}
\label{eq:MaxSymCondition}
\Pi_1^{q,t}(p) = 0,
\end{equation}
and implicitly constrains the composite spectrum~\cite{Csaki:2017cep}. Under this condition, the effective potential becomes finite and fully calculable, since the divergences associated with the $\Pi_1$ terms are absent. This significantly reduces the tuning required to stabilize the electroweak scale. The Maximal Symmetry parameter space therefore represents a distinguished region within the general MCHM framework. In the subsequent section, we ask whether this condition, usually motivated by naturalness, can also be recovered from the constraint of quantum entanglement maximization.

\section{Entanglement Constraints in the Fermion Sector}
\label{Sec:EC}

\par The unitary evolution of a closed quantum system implies that its global von Neumann entropy is conserved. In contrast, the entanglement entropy of a subsystem typically increases. This phenomenon is analogous to thermodynamic entropy production and represents the delocalization of quantum information into internal degrees of freedom. Within the framework of high-energy physics, $2 \to 2$ scattering processes constitute the fundamental microscopic mechanism that governs the generation of quantum correlations. Therefore, the capacity of scattering amplitudes to generate entanglement provides a direct quantitative measure of the correlations inherent in the theory.

\par We now apply the principle of entanglement extremization to the electroweak symmetry breaking sector. This approach is motivated by the observation that entanglement suppression characterizes low-energy QCD~\cite{Beane_2019}. Specifically, we impose the requirement that the scattering of Higgs bosons into top quarks generates a maximally entangled final state, and examine how this information-theoretic condition constrains the parameter space of the effective field theory derived in Sec.~\ref{sec:CHM}. By analyzing the effective Lagrangian under this constraint, we aim to identify the form-factor structures that are compatible with the maximization of quantum correlations.

\par To characterize the low-energy phenomenology, we adopt the EFT framework. We expand the Goldstone matrix $\Sigma^{\prime}$ in powers of the Higgs doublet
\begin{equation}
H = (h_1 + i h_2,\; h_4 + i h_3).
\end{equation}
By truncating this expansion at second order and restricting the analysis to the electrically neutral components, we identify the scalar Higgs boson $h_4$ and the pseudoscalar Goldstone boson $h_3$ as the relevant degrees of freedom. We then substitute these expansions into the effective Lagrangian defined in Eq.~(\ref{eq:Lagrangian}). This procedure yields the leading-order effective interactions between the Higgs sector and the top quark. Focusing specifically on the terms that contribute to the $hh \to t\bar{t}$ scattering process, the effective Lagrangian contains the following interactions~\cite{marzocca2012general,Panico_2016}:
\begin{align}
\label{eq:Leff_interaction}
\mathcal{L}_{\text{eff}} \supset\;&
\frac{\sqrt{2}\, M_1^t(p)}{f}
\bigl[
i h_3 (\bar{t}_R t_L - \bar{t}_L t_R)
+ h_4 (\bar{t}_R t_L + \bar{t}_L t_R)
\bigr]
\nonumber\\[2pt]
&-
\frac{h_3^2 + h_4^2}{f^2}
\bigl[
\Pi_{1}^{q}\, \bar{t}_L \slashed{p} t_L
- 2 \Pi_{1}^{t}\, \bar{t}_R \slashed{p} t_R
\bigr].
\end{align}
Here, the first line arises from the effective Yukawa couplings proportional to the form factor $M_1^t$, while the second line is governed by the kinetic form factors $\Pi_1^{q,t}$. We have omitted the interactions involving the bottom quark, since they do not contribute to the $hh \to t\bar{t}$ amplitude at tree level.

\par
\begin{figure}[htbp]
    \centering
    \tikzset{
        scalar/.style={draw=black, dashed, thick},
        fermion/.style={
            draw=black,
            thick,
            postaction={decorate},
            decoration={
                markings,
                mark=at position 0.6 with {\arrow{Stealth[length=2.5mm,width=1.5mm]}}
            }
        },
        dot/.style={fill=black, circle, minimum size=5pt, inner sep=0pt}
    }
    \def\w{1.5}
    \def\h{1.6}
    \def\v{0.8}
    \begin{tikzpicture}
        \begin{scope}[shift={(0,0)}]
            \coordinate (in_top) at (-\w,\h);
            \coordinate (in_bot) at (-\w,-\h);
            \coordinate (v_top)  at (0,\v);
            \coordinate (v_bot)  at (0,-\v);
            \coordinate (out_top) at (\w,\h);
            \coordinate (out_bot) at (\w,-\h);

            \draw[scalar] (in_top) -- (v_top);
            \draw[scalar] (in_bot) -- (v_bot);
            \draw[thick]  (v_top) -- (v_bot);
            \draw[fermion] (v_top) -- (out_top);
            \draw[fermion] (out_bot) -- (v_bot);

            \node[left]  at (in_top)  {\Large $h$};
            \node[left]  at (in_bot)  {\Large $h$};
            \node[right] at (out_top) {\Large $t$};
            \node[right] at (out_bot) {\Large $\bar t$};
        \end{scope}

        \begin{scope}[shift={(4.5,0)}]
            \coordinate (in_top) at (-\w,\h);
            \coordinate (in_bot) at (-\w,-\h);
            \coordinate (v_top)  at (0,\v);
            \coordinate (v_bot)  at (0,-\v);
            \coordinate (out_top) at (\w,\h);
            \coordinate (out_bot) at (\w,-\h);

            \draw[scalar] (in_top) -- (v_top);
            \draw[scalar] (in_bot) -- (v_bot);
            \draw[thick]  (v_top) -- (v_bot);
            \draw[fermion] (out_bot) -- (v_top);
            \draw[fermion] (v_bot) -- (out_top);

            \node[left]  at (in_top)  {\Large $h$};
            \node[left]  at (in_bot)  {\Large $h$};
            \node[right] at (out_top) {\Large $t$};
            \node[right] at (out_bot) {\Large $\bar t$};
        \end{scope}

        \begin{scope}[shift={(9,0)}]
            \coordinate (in_top) at (-\w,\h);
            \coordinate (in_bot) at (-\w,-\h);
            \coordinate (center) at (0,0);
            \coordinate (out_top) at (\w,\h);
            \coordinate (out_bot) at (\w,-\h);

            \draw[scalar] (in_top) -- (center);
            \draw[scalar] (in_bot) -- (center);
            \draw[fermion] (center) -- (out_top);
            \draw[fermion] (out_bot) -- (center);
            \node[dot] at (center) {};

            \node[left]  at (in_top)  {\Large $h$};
            \node[left]  at (in_bot)  {\Large $h$};
            \node[right] at (out_top) {\Large $t$};
            \node[right] at (out_bot) {\Large $\bar t$};
        \end{scope}
    \end{tikzpicture}
    \caption{Tree-level Feynman diagrams for the process $hh \to t\bar{t}$. From left to right: the $t$-channel diagram, the $u$-channel diagram involving top-quark exchange, and the four-point contact interaction vertex.}
    \label{fig:hh_ttbar}
\end{figure}
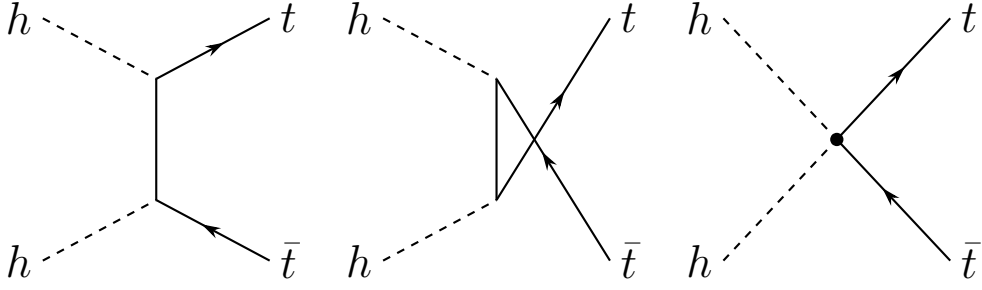

 We analyze the annihilation process of Higgs bosons into a top-antitop pair, $hh \to t\bar{t}$. To quantify the entanglement generated in the process, we prepare the initial states in the interaction basis
\begin{equation}
\ket{h_3h_3},\qquad
\ket{h_3h_4},\qquad
\ket{h_4h_3},\qquad
\ket{h_4h_4},
\end{equation}
and project the resulting scattering amplitude onto the $t\bar{t}$ helicity subspace spanned by
\begin{equation}
\ket{R\bar{R}},\qquad
\ket{R\bar{L}},\qquad
\ket{L\bar{R}},\qquad
\ket{L\bar{L}}.
\end{equation}
The relevant tree-level Feynman diagrams are shown in Fig.~\ref{fig:hh_ttbar}.

\par In the high-energy limit considered here, chirality is approximately conserved and only the opposite-helicity amplitudes contribute. Therefore, the amplitudes with identical fermion helicities vanish,
\begin{equation}
\mathcal{M}_{LL} = \mathcal{M}_{RR} = 0,
\end{equation}
so that the final state reduces to
\begin{equation}
\ket{\Psi}
=
\frac{1}{\sqrt{\mathcal N}}
\left(
\mathcal{M}_{LR}\ket{01}
+
\mathcal{M}_{RL}\ket{10}
\right),
\quad
\mathcal N = |\mathcal M_{LR}|^2 + |\mathcal M_{RL}|^2.
\end{equation}
Substituting this form into Eq.~(\ref{eq:concurrence}), the concurrence becomes
\begin{equation}
\label{eq:concurrence_reduced}
\Delta(\ket{\Psi})
=
\frac{2|\mathcal M_{LR}\mathcal M_{RL}|}
{|\mathcal M_{LR}|^2 + |\mathcal M_{RL}|^2}.
\end{equation}
The upper bound is saturated if and only if
\begin{equation}
\label{eq:ME_condition}
\Delta = 1
\quad \Longleftrightarrow \quad
|\mathcal M_{LR}| = |\mathcal M_{RL}|.
\end{equation}
Hence, within the present two-dimensional helicity subspace, maximal entanglement is exactly equivalent to the equality in magnitude of the two opposite-helicity amplitudes.

\par For initial states with distinct interaction eigenstates (\emph{i.e.}, $h_3h_4$ and $h_4h_3$), the four-point contact interaction is absent because of CP symmetry. The tree-level scattering amplitude is therefore determined solely by the $t$-channel and $u$-channel diagrams. We employ the spinor-helicity formalism to derive compact analytical expressions for these amplitudes~\cite{Elvang:2013cua}. The resulting non-vanishing helicity amplitudes are
\begin{align}
M_{h_3h_4 \to t_L \bar{t}_R}
=
M_{h_4h_3 \to t_L \bar{t}_R}
&=
\frac{2i |M^t_1|^2}
{f^2 \sqrt{(\Pi_0^q + \Pi_1^q)(\Pi_0^t - \Pi_1^t)}}
\left( \frac{1}{t} - \frac{1}{u} \right)
\asbk{324},
\\[2pt]
M_{h_3h_4 \to t_R \bar{t}_L}
=
M_{h_4h_3 \to t_R \bar{t}_L}
&=
\frac{2i |M^t_1|^2}
{f^2 \sqrt{(\Pi_0^q + \Pi_1^q)(\Pi_0^t - \Pi_1^t)}}
\left( \frac{1}{u} - \frac{1}{t} \right)
\sabk{324}.
\end{align}

\par The spinor structures associated with the two non-vanishing helicity amplitudes are related by complex conjugation. This property ensures that the amplitudes possess identical magnitudes,
\begin{equation}
|M_{h_3h_4 \to t_L \bar{t}_R}|
=
|M_{h_3h_4 \to t_R \bar{t}_L}|,
\end{equation}
and similarly for the $h_4h_3$ channel. Consequently, Eq.~(\ref{eq:ME_condition}) is automatically satisfied, so that the final state is maximally entangled and characterized by a concurrence of unity across the entire kinematic phase space. Within the present high-energy helicity-projected setup, the scattering of distinct Higgs interaction eigenstates, $h_3$ and $h_4$, therefore produces a maximally entangled top-antitop configuration.

\par Next, we examine initial states composed of identical Higgs bosons. We first calculate the contributions arising specifically from the $t$- and $u$-channel exchanges. These pole-mediated amplitudes take the form
\begin{align}
M^{\text{tree}}_{h_3h_3 \to t_L \bar{t}_R}
=
M^{\text{tree}}_{h_4h_4 \to t_L \bar{t}_R}
&=
\frac{2 |M^t_1|^2}
{f^2 \sqrt{(\Pi_0^q + \Pi_1^q)(\Pi_0^t - \Pi_1^t)}}
\left( \frac{1}{t} + \frac{1}{u} \right)
\asbk{324},
\\[2pt]
M^{\text{tree}}_{h_3h_3 \to t_R \bar{t}_L}
=
M^{\text{tree}}_{h_4h_4 \to t_R \bar{t}_L}
&=
\frac{2 |M^t_1|^2}
{f^2 \sqrt{(\Pi_0^q + \Pi_1^q)(\Pi_0^t - \Pi_1^t)}}
\left( \frac{1}{t} + \frac{1}{u} \right)
\sabk{324}.
\end{align}

\par In the absence of the contact interaction, the scattering of identical Higgs bosons also produces a maximally entangled final state, since the two spinor structures are again related by complex conjugation and therefore have equal magnitude. However, the inclusion of the four-point contact term modifies this picture. Its contributions to the helicity amplitudes are
\begin{align}
M^{\text{contact}}_{h_3h_3 \to t_L \bar{t}_R}
=
M^{\text{contact}}_{h_4h_4 \to t_L \bar{t}_R}
&=
\frac{\Pi_1^q}{f^2 (\Pi_0^q + \Pi_1^q)} \asbk{3p4},
\\[2pt]
M^{\text{contact}}_{h_3h_3 \to t_R \bar{t}_L}
=
M^{\text{contact}}_{h_4h_4 \to t_R \bar{t}_L}
&=
-\frac{2 \Pi_1^t}{f^2 (\Pi_0^t - \Pi_1^t)} \sabk{3p4}.
\end{align}

\par The total scattering amplitude consists of the coherent sum of the pole and contact contributions. Writing
\begin{align}
\mathcal M_{LR} &= P\asbk{324} + C_L\asbk{3p4},
\\
\mathcal M_{RL} &= P\sabk{324} + C_R\sabk{3p4},
\end{align}
where
\begin{align*}
P &=
\frac{2 |M^t_1|^2}
{f^2 \sqrt{(\Pi_0^q + \Pi_1^q)(\Pi_0^t - \Pi_1^t)}}
\left( \frac{1}{t} + \frac{1}{u} \right), \\
C_L &= \frac{\Pi_1^q}{f^2 (\Pi_0^q + \Pi_1^q)},
\qquad
C_R = -\frac{2 \Pi_1^t}{f^2 (\Pi_0^t - \Pi_1^t)},
\end{align*}
the exact condition for maximal entanglement is
\begin{equation}
\label{eq:exact_condition}
\bigl| P\asbk{324} + C_L\asbk{3p4} \bigr|
=
\bigl| P\sabk{324} + C_R\sabk{3p4} \bigr|.
\end{equation}
In general, Eq.~(\ref{eq:exact_condition}) may be satisfied through nontrivial interference between the pole and contact terms. Nevertheless, since the pole contribution already saturates the maximal entanglement condition by itself, a natural and kinematically robust sufficient condition for the full amplitude to preserve this property is that the contact terms do not spoil the equality of the two opposite-helicity amplitudes. This is achieved if the left- and right-handed contact contributions have equal magnitude. In the tree-level real-parameter setup considered here, the form factors are real away from resonance poles. We focus on the aligned real branch in which the contact terms preserve the same conjugate structure as the pole contribution, namely $C_L=C_R$. This gives
\begin{equation}
\label{eq:max_ent_condition}
\frac{\Pi_{1}^{q}}{\Pi_{0}^{q} + \Pi_{1}^{q}}
=
-\frac{2 \Pi_{1}^{t}}{\Pi_{0}^{t} - \Pi_{1}^{t}}.
\end{equation}

\par Eq.~\eqref{eq:max_ent_condition} represents a functional relation that must hold across the momentum dependence of the effective theory if maximal entanglement is to be interpreted as a structural property rather than an accidental feature at a special kinematic point. Since the form factors $\Pi_{0,1}^{q,t}$ are nontrivial functions of momentum, satisfying this identity generically requires a nontrivial relation between the underlying masses and mixings. Substituting the explicit expressions for the form factors, one finds
\begin{align}
\label{eq:form_factor_constraint}
\frac{\epsilon_{qQ}^2 (M_S^2 - p^2) + \epsilon_{qS}^2 (p^2 - M_Q^2)}
     {(M_S^2 - p^2)\bigl(p^2 - M_Q^2 + 2\epsilon_{qQ}^2\bigr)}
=
-\frac{\epsilon_{tQ}^2 (M_S^2 - p^2) + \epsilon_{tS}^2 (p^2 - M_Q^2)}
      {(M_Q^2 - p^2)\bigl(p^2 - M_S^2 + \epsilon_{tS}^2\bigr)}.
\end{align}

\par A particularly simple symmetric branch of solutions to Eq.~\eqref{eq:form_factor_constraint} is obtained by requiring the numerators on both sides to vanish identically. This yields
\begin{equation}
\label{eq:alg_conditions_1}
M_Q^2 = M_S^2,
\qquad
\epsilon_{tQ}^2 = \epsilon_{tS}^2,
\qquad
\epsilon_{qQ}^2 = \epsilon_{qS}^2.
\end{equation}
This branch will be the focus of the following analysis, since it admits a direct interpretation in terms of enhanced symmetry and a nontrivial phenomenological selection through the behavior of the top Yukawa coupling.

\par
To make this interpretation explicit, it is useful to separate the information contained in Eq.~\eqref{eq:alg_conditions_1} into two parts: a convention for the relative signs of the mixing parameters and a genuine mass relation.  In a
real-parameter convention, the last two relations in
Eq.~\eqref{eq:alg_conditions_1} imply that each pair of mixings can differ at most by a relative sign. These signs are not physical by themselves, since they can be shifted between the mixing terms and the Dirac mass terms of the composite fermions through chiral phase redefinitions of the fermion partners. Such field redefinitions leave the physical pole masses and the condition $M_Q^2=M_S^2$ unchanged. After choosing the aligned basis, the remaining sign information is carried by the relative sign between $M_Q$ and $M_S$. We may therefore work in the aligned basis
\begin{equation}
    \epsilon_{tQ}=\epsilon_{tS},\qquad
\epsilon_{qQ}=\epsilon_{qS}.
\end{equation}
Equivalently, with
\begin{equation}
    c_{\pm L}=\frac{\epsilon_{qQ}\pm\epsilon_{qS}}{\sqrt{2}},
\qquad
c_{\pm R}=\frac{\epsilon_{tQ}\pm\epsilon_{tS}}{2},
\end{equation}
this choice gives \(c_{-L}=c_{-R}=0\).  The mixing-sector conditions in Eq.~\eqref{eq:alg_conditions_1} are then automatically satisfied, and the only remaining nontrivial constraint is the mass relation
\begin{equation}
    (M_Q+M_S)(M_Q-M_S)=0 .    
\end{equation}
There are therefore two algebraic branches:
\begin{itemize}
    \item \(M_Q+M_S=0\), corresponding to the twisted mass relation of the
    maximal-symmetry branch;
    \item \(M_Q-M_S=0\), corresponding to the symmetric branch.
\end{itemize}

The second branch is, however, physically trivial for the present analysis.
Indeed, the form factor controlling the effective top Yukawa coupling is
proportional to
\begin{equation}
\label{eq:M1t_numerator}
M_1^t(p)
\propto
\epsilon_{qQ}\epsilon_{tQ} M_Q (p^2-M_S^2)
-
\epsilon_{qS}\epsilon_{tS} M_S (p^2-M_Q^2).
\end{equation}
In the aligned basis, the symmetric branch \(M_Q=M_S\) makes the two
terms cancel identically, so that
\begin{equation}
    M_1^t(p)=0 .    
\end{equation}
This simultaneously removes the effective top Yukawa coupling, gives a
massless top quark, and eliminates the \(hh\to t\bar t\) amplitude entering
the concurrence analysis. Hence this algebraic branch is not a viable
composite-Higgs solution. Within this aligned symmetric branch, the only physically nontrivial branch is therefore
\begin{equation}
    M_Q+M_S=0,
\end{equation}
which is precisely the maximal-symmetry condition.

\par In addition to the numerator-vanishing branch discussed above, Eq.~\eqref{eq:max_ent_condition} also admits a distinct matching branch. This second class of solutions is characterized by
\begin{equation}
\label{eq:redefined_factor}
\Pi_1 = \Pi_1^q = -2\Pi_1^t,
\qquad
\Pi_0 = \Pi_0^t - \Pi_1^t = \Pi_0^q + \Pi_1^q .
\end{equation}
Here, $\Pi_0$ and $\Pi_1$ are introduced only as shorthand notations for the corresponding form-factor combinations. These relations define a generalized $Z_2$-matching branch, in which the left- and right-handed top sectors contribute to the contact amplitudes with the same effective coefficient. This branch provides a second structural way of satisfying the maximal-entanglement condition, distinct from the numerator-vanishing Maximal Symmetry branch obtained above. 

\par Requiring Eq.~\eqref{eq:redefined_factor} to hold as a functional identity in $p^2$, and focusing on the nontrivial branch with nonvanishing residues, the matching of pole structures and residues gives the following conditions. Here we have chosen a common normalization for the two chiral sectors, $\lambda_L=\lambda_R$, one obtains
\begin{equation}
M_Q^2=M_S^2,
\qquad
2\epsilon_{qQ}^2=\epsilon_{tS}^2,
\qquad
\epsilon_{qQ}^2-\epsilon_{qS}^2
=
\epsilon_{tS}^2-\epsilon_{tQ}^2 .
\end{equation}
Thus, in the two-resonance parametrization used here, the generalized $Z_2$ branch corresponds to a nontrivial left-right matching of the microscopic masses and mixings. This provides an additional symmetry structure compatible with maximal entanglement in the fermionic effective theory.

\par On the generalized $Z_2$ branch, the effective Lagrangian may be reorganized in terms of the unified form factors in Eq.~\eqref{eq:redefined_factor}. The Higgs-dependent top-sector interactions then take the form
\begin{align}
\mathcal{L}_{\text{eff}}
=&\;
\frac{\sqrt{2} M_t}{f}
\left[
i h_3 (\bar{t}_R t_L - \bar{t}_L t_R)
+ h_4 (\bar{t}_R t_L + \bar{t}_L t_R)
\right]
\nonumber\\[2pt]
&+
\Pi_{0}
\left(
\bar{b}_L \slashed{p} b_L
+
\bar{t}_L \slashed{p} t_L
+
\bar{t}_R \slashed{p} t_R
\right)
\nonumber\\[2pt]
&-
\frac{h_3^2 + h_4^2}{f^2}\,
\Pi_1
\left(
\bar{t}_L \slashed{p} t_L
+
\bar{t}_R \slashed{p} t_R
\right).
\end{align}
This form makes manifest an emergent $Z_2$ exchange structure between the left- and right-handed top sectors. In particular, the Higgs-dependent part of the Lagrangian is invariant under the discrete transformation
\begin{equation}
\label{eq:Z2_transform}
t_L \leftrightarrow t_R,
\qquad
h_3 \to -h_3,
\qquad
h_4 \to h_4,
\end{equation}
under which the two chiral sectors are exchanged while the CP-odd scalar $h_3$ changes sign. The generalized $Z_2$ branch therefore represents a distinct left-right matching solution of the maximal-entanglement condition. This generalized $Z_2$ branch should be distinguished from the Maximal Symmetry branch. Although it satisfies the maximal-entanglement matching condition, it does not by itself ensure the finiteness of the Higgs potential or reduce electroweak tuning. Therefore, it is not a phenomenologically viable naturalness branch of the CHM, but rather an additional structural solution of the entanglement constraint. 

\par It is useful to contrast these fermionic matching conditions with the gauge sector. There, the Higgs interactions with vector bosons are fixed by the covariant derivative and therefore much more rigid than the fermionic interactions governed by the form factors $\Pi^{q,t}_{0,1}$ and $M_1^t$. In the transverse gauge-boson helicity subspace, the CP-even gauge interaction implies equal magnitudes for the conjugate-helicity amplitudes, $|\mathcal A_{++}| = |\mathcal A_{--}|$. Hence the corresponding two-state helicity configuration is automatically maximally entangled. This should be understood not as an additional dynamical selection, but as a consequence of the pre-existing gauge symmetry structure. Similarly, after the heavy vector and axial resonances are included, the gauge contribution to the Higgs potential is rendered finite by the usual high-energy consistency conditions, such as the Weinberg sum rules. The gauge sector therefore serves as a reference case in which both maximal entanglement in the relevant helicity subspace and the finiteness of the Higgs potential are automatic. Further details of the entanglement calculation in this sector are provided in App.~\ref{app:gauge-sector-comparison}.

\par These results show that information-theoretic constraints place nontrivial restrictions on the fermionic EFT. Unlike the gauge sector, the top sector does not generically possess the amplitude matching required for maximal entanglement, nor does it automatically eliminate the Higgs-dependent form factor $\Pi_1$. The maximal-entanglement condition therefore identifies two simple structural ways of satisfying the constraint. The first is the numerator-vanishing branch, whose nonzero-amplitude realization corresponds to Maximal Symmetry and gives a finite Higgs potential. The second is the generalized $Z_2$-matching branch, in which the left- and right-handed top sectors obey a nontrivial matching relation among the relevant form-factor combinations. In this sense, entanglement extremization provides a useful organizing principle for identifying symmetry structures relevant to EWSB in Composite Higgs models.

\section{Conclusion}
\label{Sec:Concl}

\par 

\label{Sec:Concl}

\par In this work, we have investigated the interplay between quantum information constraints and the symmetry structures of the electroweak sector. By analyzing the process $hh \to t\bar{t}$ within the MCHM, we proposed that the requirement of maximal entanglement captures an important structural feature of the underlying strong dynamics and may be viewed as a manifestation of entanglement extremization. This perspective resonates with recent observations that entanglement suppression characterizes low-energy QCD~\cite{Beane_2019}. In the present case, the condition of maximal entanglement imposes nontrivial restrictions on the fermionic effective theory and identifies highly constrained structural solutions for the Higgs-dependent top-sector interactions. The comparison with the gauge sector further clarifies the special role of the top sector. In the gauge sector, the relevant helicity-amplitude matching is already enforced by the covariant-derivative structure, and the finiteness of the Higgs potential follows from the usual resonance completion and high-energy sum rules. The fermion sector is therefore the genuinely nontrivial arena in which entanglement maximization acts as a selection principle.

\par Our analysis shows that the maximal-entanglement condition admits two simple structural realizations in the fermionic effective theory. The first is the numerator-vanishing branch. After requiring a nonzero $hh\to t\bar t$ final state, this branch selects the Maximal Symmetry condition, corresponding to the vanishing of the Higgs-dependent form factors $\Pi_1^{q,t}$ and the finiteness of the Higgs potential. The second is a generalized $Z_2$-matching branch, in which the left- and right-handed top sectors obey a nontrivial relation among the relevant form-factor combinations. These two branches provide distinct symmetry-based ways of satisfying the maximal-entanglement condition.

\par Two aspects of this result are particularly noteworthy. First, enforcing the entanglement condition across the momentum dependence of the effective theory leads to nontrivial relations among the underlying masses and mixings of the composite resonances. Second, the resulting solutions connect the entanglement structure of the scattering amplitudes to symmetry patterns in the top sector: the Maximal Symmetry branch provides the finite-potential realization relevant for electroweak naturalness, while the generalized $Z_2$ branch reveals an additional left-right matching structure compatible with maximal entanglement.

\par Taken together, these results support the idea that entanglement extremization may serve as a useful organizing principle for strong dynamics. While low-energy QCD exhibits entanglement suppression, our findings indicate that the composite sector governing electroweak symmetry breaking is naturally associated with entanglement maximization. This contrast suggests that strong dynamics may be driven toward extrema of quantum correlation. A natural direction for future work is to extend this framework to other strongly coupled sectors and to clarify more systematically the relation between entanglement structure, emergent symmetry, and naturalness.

\section*{Acknowledgments}

This work was supported by the National Natural Science Foundation of China (grant Nos. 12450006, E514660101), Innovation
 Program for Quantum Science and Technology
(Grant No. 2021ZD0303205), Youth Innovation
 Promotion Association (Grant No. 2023474), and National
 Key Research and Development Program of China
(grant no. 2018YFA0306600). T.M. is partly supported by Chinese Academy of Sciences Pioneer Initiative "Talent Introduction Plan" (grant No.\,E4ER6601A2), the Fundamental Research Funds for the Central Universities (grant No.\,E4EQ6602X2), and the National Natural Science Foundation of China (grant No.\,E514660101).

\vspace{1.0cm}

\appendix
\section{Form factors}

\par For completeness, we collect here the explicit expressions for the form factors entering Eq.~(\ref{eq:Leff_interaction}).

\begin{align*}
\frac{\Pi_0^{q,t}}{\lambda_{L,R}^2 f^2}
=&\;
1
+\frac{(c_{-L,R}^2 + c_{+L,R}^2)(M_Q^2 + M_S^2 - 2p^2)}
{2(p^2 - M_S^2)(M_Q^2 - p^2)}
\\[2pt]
&\;
+\frac{c_{-L,R}c_{+L,R}(M_S + M_Q)(M_S - M_Q)}
{(p^2 - M_S^2)(M_Q^2 - p^2)} .
\end{align*}

\begin{align*}
\frac{\Pi_1^{q,t}}{\lambda_{L,R}^2 f^2}
=&\;
\frac{c_{+L,R}c_{-L,R}(M_Q^2 + M_S^2 - 2p^2)}
{(p^2 - M_S^2)(M_Q^2 - p^2)}
\\[2pt]
&\;
+\frac{(c_{+L,R}^2 + c_{-L,R}^2)(M_S - M_Q)(M_S + M_Q)}
{2(p^2 - M_S^2)(M_Q^2 - p^2)} .
\end{align*}

\begin{align*}
\frac{M_1^t}{\lambda_L \lambda_R f^2}
=&\;
\frac{M_Q^2 M_S (c_{-L} - c_{+L})(c_{-R} - c_{+R})}
{2(p^2 - M_Q^2)(p^2 - M_S^2)}
-\frac{M_S^2 M_Q (c_{-L} + c_{+L})(c_{-R} + c_{+R})}
{2(p^2 - M_Q^2)(p^2 - M_S^2)}
\\[2pt]
&\;
+\frac{M_Q (c_{-L} + c_{+L})(c_{-R} + c_{+R})\,p^2}
{2(p^2 - M_Q^2)(p^2 - M_S^2)}
-\frac{M_S (c_{-L} - c_{+L})(c_{-R} - c_{+R})\,p^2}
{2(p^2 - M_Q^2)(p^2 - M_S^2)} .
\end{align*}

\par Here, we have factored out the overall coupling strengths $\lambda_{L,R}$, while the structural mixing parameters $\epsilon$ are encoded in the linear combinations
\begin{equation}
c_{\pm R} = \frac{\epsilon_{tQ} \pm \epsilon_{tS}}{2},
\qquad
c_{\pm L} = \frac{\epsilon_{qQ} \pm \epsilon_{qS}}{\sqrt{2}} .
\end{equation}
For $\Pi_{0,1}^q$ one should read $c_{\pm L}$ and $\lambda_L$, while for $\Pi_{0,1}^t$ one should read $c_{\pm R}$ and $\lambda_R$.

\section{Gauge-sector comparison}
\label{app:gauge-sector-comparison}

\par For completeness, we provide here the gauge-sector comparison briefly discussed at the end of Sec.~\ref{Sec:EC}. The purpose of this comparison is not to introduce an additional selection principle, but rather to clarify why the fermion sector is the genuinely nontrivial arena for the entanglement-based constraint. In the fermion sector, the relevant Higgs-dependent interactions are controlled by the form factors $\Pi^{q,t}_{0,1}$ and $M_1^t$. Consequently, the relative magnitude of the two opposite-helicity amplitudes in $hh\to t\bar t$ is not fixed by symmetry alone. This is why imposing maximal entanglement leads to the nontrivial condition in Eq.~\eqref{eq:max_ent_condition}.

\par By contrast, the gauge sector is much more rigid. The interactions between the Higgs field and the electroweak gauge bosons are fixed by the covariant derivative structure of the nonlinear sigma model. As a result, the corresponding helicity amplitudes are already strongly constrained by gauge invariance and CP. To make this point explicit, let us focus on the transverse helicity subspace of the final-state gauge bosons. In this subspace, the relevant two-state basis is
\begin{align}
    \left\{ \ket{++},\ket{--} \right\}.
    \label{eq:gauge_transverse_basis}
\end{align}

\par The corresponding projected amplitudes will be denoted by $\mathcal A_{++}$ and $\mathcal A_{--}$. For the CP-even gauge interaction generated by the covariant derivative, the two conjugate-helicity amplitudes are related by complex conjugation up to an overall phase. Therefore they have equal magnitude,
\begin{align}
    \left|\mathcal A_{++}\right|
    =
    \left|\mathcal A_{--}\right| .
    \label{eq:gauge_helicity_equality}
\end{align}
\par The projected final state in this transverse helicity subspace may then be written as
\begin{align}
    \ket{\Psi_V}
    =
    \frac{1}{\sqrt{\mathcal N_V}}
    \left(
        \mathcal A_{++}\ket{++}
        +
        \mathcal A_{--}\ket{--}
    \right),
    \qquad
    \mathcal N_V
    =
    \left|\mathcal A_{++}\right|^2
    +
    \left|\mathcal A_{--}\right|^2 .
    \label{eq:gauge_projected_state}
\end{align}
\par Using the same concurrence formula introduced in Sec.~\ref{sec:Ent}, the entanglement of this two-state helicity configuration is
\begin{align}
    \Delta_V
    =
    \frac{
        2\left|\mathcal A_{++}\mathcal A_{--}\right|
    }{
        \left|\mathcal A_{++}\right|^2
        +
        \left|\mathcal A_{--}\right|^2
    } .
    \label{eq:gauge_concurrence}
\end{align}
\par Substituting Eq.~\eqref{eq:gauge_helicity_equality} into
Eq.~\eqref{eq:gauge_concurrence}, one obtains
\begin{align}
    \Delta_V = 1 .
    \label{eq:gauge_maximal_entanglement}
\end{align}
\par Thus, within the relevant transverse helicity subspace, the gauge sector automatically produces a maximally entangled two-state configuration. The key point is that this result does not require an additional dynamical condition analogous to Eq.~\eqref{eq:max_ent_condition}. Instead, the amplitude matching required for maximal entanglement is already enforced by the CP-even covariant-derivative interaction and the underlying gauge symmetry.

\par This behavior also parallels the structure of the Higgs potential in the gauge sector of composite Higgs models. The low-energy description containing only the light electroweak gauge bosons is incomplete in the ultraviolet. Once the heavy
vector and axial resonances associated with the strong sector are included, the relevant current correlators obey the usual high-energy consistency conditions, which may be expressed through Weinberg sum rules~\cite{Csaki:2017jby}. These
relations ensure the cancellation of the leading UV-sensitive contributions to the gauge part of the Higgs potential. Schematically, the required high-energy falloff may be represented by the cancellation of the leading terms in the
difference of vector and axial current correlators,
\begin{align}
    \Pi_V(p^2)-\Pi_A(p^2)
    \;\xrightarrow{p^2\to\infty}\;
    \mathcal O\!\left(\frac{1}{p^4}\right),
    \label{eq:gauge_sum_rule_falloff}
\end{align}
so that the corresponding gauge contribution to the Higgs potential is finite and calculable after resonance completion.

\par The gauge sector therefore serves as a useful reference case. Both the maximal entanglement of the relevant transverse helicity configuration and the finiteness of the Higgs potential follow from structures that are already present: gauge invariance, CP-even covariant-derivative interactions, and the high-energy behavior of the completed resonance theory. The top sector is qualitatively different. There, the Higgs-dependent form factors are not fixed by gauge symmetry alone, and the equality of the two opposite-helicity amplitudes is not automatic. It is precisely for this reason that the maximal-entanglement condition becomes a nontrivial selection principle in the fermion sector.
\vspace{1.0cm}

\bibliography{AMP}
\bibliographystyle{apsrev} 

\end{document}